\documentclass[twocolumn]{aastex701}
\usepackage{graphicx}
\usepackage{subcaption}
\usepackage{siunitx}
\usepackage{amssymb}
\newcommand{\scinot}[1]{\times 10^{#1}}
\newcommand{\msun}{M_{\odot}}

\begin{document}
\linenumbers

\title{Comparative Bayesian SED Fitting of PEARLSDG}

\author[orcid=0009-0004-6194-1899,sname='Bayraktar']{Barış Bayraktar}
\affiliation{School of Earth and Space Exploration, Arizona State University,
Tempe, AZ 85287-6004, USA}
\email[]{bbayrakt@asu.edu}

\author[orcid=0000-0001-6650-2853,sname='Timothy']{Timothy Carleton}
\affiliation{School of Earth and Space Exploration, Arizona State University,
Tempe, AZ 85287-6004, USA}
\email[]{tmcarlet@asu.edu}

\author[orcid=0000-0001-8156-6281]{Rogier A.\ Windhorst}
\affiliation{School of Earth and Space Exploration, Arizona State University,
Tempe, AZ 85287-6004, USA}
\email{Rogier.Windhorst@asu.edu}

\author[orcid=0000-0002-9895-5758]{S.\ P.\ Willner}
\affiliation{Center for Astrophysics \textbar\ Harvard \& Smithsonian, 60 Garden Street,
Cambridge, MA 02138, USA}
\email{swillner@cfa.harvard.edu}

\author[orcid=0000-0001-9262-9997]{Christopher N.\ A.\ Willmer}
\affiliation{Steward Observatory, University of Arizona, 933 N Cherry Ave,
Tucson, AZ 85721-0009, USA}
\email{cnaw@arizona.edu}

\begin{abstract}
The initial distance to PEARLSDG estimated from the Tip of the Red Giant Branch
suggested it was an exotic isolated quiescent dwarf galaxy. We combine recent and
archival Hectospec spectroscopy to place it at $z = 0.02843\pm0.00012$
($D \approx 124$\,Mpc) within a galaxy group, revising the distance from 30\,Mpc
to $\sim$124\,Mpc. We then carry out {\sc Prospector} SED fitting using parametric
and non-parametric star-formation histories sampled with \texttt{dynesty},
\texttt{nautilus}, and \texttt{emcee}, recovering metallicity
$\log(Z/Z_\odot) = -0.44^{+0.35}_{-0.06}$, stellar mass
$\log_{10}(M_*/\msun) = 9.25^{+0.02}_{-3.73}$, and dust attenuation
$\hat{\tau}_V = 0.67^{+0.02}_{-0.05}$. The updated metallicity places PEARLSDG
squarely on the standard mass--metallicity relation, resolving its former outlier
status, with its quenched star-formation history consistent with environmental
quenching in a group setting.
\end{abstract}

\section{Introduction} \label{sec:intro}

PEARLSDG was initially classified \citep{carleton_pearls_2024} as a quiescent,
isolated dwarf at a Tip of the Red Giant Branch (TRGB) distance of 30~Mpc. New
spectroscopy \citep{carleton_new_2024} revealed $z = 0.02843\pm0.00012$,
corresponding to a luminosity distance of $\sim$124~Mpc. At this distance,
PEARLSDG appears associated with a galaxy group at $z = 0.0273$ comprising at
least five SDSS members with stellar masses $10^{9}$--$10^{10}~\msun$, projected
separations of 128--548~kpc, and line-of-sight velocity offsets of
68--565~km~s$^{-1}$. Accurately constraining PEARLSDG's stellar population
parameters under this group association requires robust SED modeling incorporating
the new spectroscopic data. We reanalyze PEARLSDG (DOI: 10.17909/09e0-
ks54) using {\sc Prospector} with both
parametric and non-parametric star-formation history (SFH) models.
Distances assume $H_0=70$~km~s$^{-1}$~Mpc$^{-1}$.

\section{Observations} \label{sec:obs}

All spectroscopy came from the Hectospec instrument on the MMT\null. The 2024
observations (PI Willner) reported by \cite{carleton_new_2024} were combined with
two archival exposures \citep{rines_spectroscopic_2022} taken in 2007 (3000~s and
4800~s). A combined, exposure-time-weighted spectrum yields
$z=0.02843\pm0.00012$ (statistical uncertainty only).

\section{SED Fitting} \label{sec:fit}

Fitting used {\sc Prospector} \citep{johnson_stellar_2021}, jointly modeling the
Hectospec optical spectrum and broadband photometry at fixed $z = 0.02843$. Spectral
regions affected by sky residuals or low signal-to-noise were excluded, retaining
$\sim$4000--7000\,\AA. Non-parametric SFHs divide star formation into discrete
age bins with free amplitudes, offering greater flexibility than analytic forms.
Parameter estimation employed \texttt{dynesty} (nested sampling), \texttt{nautilus}
(dynamic nested sampling), and \texttt{emcee} (affine-invariant MCMC) to validate
posteriors across algorithms.

The preferred non-parametric \texttt{emcee} model yields
$\log_{10}(M_*/\msun) = 9.25^{+0.02}_{-3.43}$, dominated by the most recent SFH
bin; earlier bins contribute negligible mass ($\log_{10}(M_*/\msun)\lesssim 5$),
consistent with a heavily quenched system. Metallicity converges to
$\log(Z/Z_\odot) = -0.45^{+0.36}_{-0.06}$ and dust attenuation to
$\hat{\tau}_V = 0.67^{+0.02}_{-0.05}$ (Figure~1, top).

The non-parametric \texttt{emcee} model best reproduces the continuum shape, broad
absorption features, and flux normalization over 4000--7000\,\AA\ (Figure~1,
middle). The \texttt{nautilus} run independently recovered consistent parameters
($\log(Z/Z_\odot) = -0.44$, $\hat{\tau}_V = 0.66$, $\log_{10}(M_*/\msun) = 9.25$),
confirming the \texttt{emcee} posteriors. \texttt{Dynesty} systematically
underestimated flux at $\gtrsim$5500\,\AA, indicating less thorough posterior
exploration in this configuration. The parametric \texttt{dynesty} fit yielded
physically implausible parameters ($\log(Z/Z_\odot)\approx -2$,
$\hat{\tau}_V\approx 8\scinot{-4}$), likely artifacts of the restrictive
functional form. The best-fit model underestimates Ca\,\textsc{ii} K, H, and
H$\delta$ absorption in the 4000--4500\,\AA\ region (Figure~1, bottom), possibly
reflecting stellar library limitations at the recovered metallicity or an
intermediate-age population not captured by the coarse SFH binning.

\section{Discussion} \label{sec:discussion}

The distance revision has direct consequences for the interpretation of PEARLSDG's star-formation history, metallicity, and quenching mechanisms.

\cite{carleton_pearls_2024} initially reported PEARLSDG as an anomalously metal-poor, isolated, quiescent dwarf based on a TRGB distance of 30\,Mpc. Under that assumption, its low inferred metallicity was difficult to reconcile with standard chemical enrichment models at its apparent mass. This tension is resolved by the updated distance: our preferred non-parametric \texttt{emcee} model yields $\log(Z/Z_\odot) \approx -0.45$, substantially higher than previously reported and consistent with the standard mass--metallicity relation for local galaxies of comparable mass.

The non-parametric \texttt{emcee} model further reveals that PEARLSDG formed the bulk of its stellar mass in the most recent SFH bin, with earlier bins contributing negligible mass. This is consistent with a scenario in which the galaxy assembled its stars over a relatively extended period before undergoing rapid quenching upon infall into the group halo, where environmental processes such as ram-pressure stripping or starvation are the most natural explanation for its present-day quiescence. The updated group membership thus renders PEARLSDG a typical environmentally quenched galaxy rather than an exotic outlier requiring non-standard internal quenching physics.

A methodological finding of this analysis is the sensitivity of the recovered parameters to both the SFH parameterization and the choice of sampler. The non-parametric \texttt{nautilus} run independently recovered parameters consistent with \texttt{emcee} to within the quoted uncertainties, lending confidence to the robustness of the preferred solution. Simultaneous modeling of broadband photometry and optical spectroscopy was central to these results: photometric fluxes anchor the overall continuum shape while spectroscopy constrains absorption features, jointly breaking the degeneracies that limit purely spectroscopic or photometric approaches.

While the statistical uncertainties on the preferred fit (Figure~1) are small, 
the quoted posteriors reflect precision only within the chosen model framework. Systematic uncertainties from SPS assumptions (IMF, binary evolution, isochrone library) are expected to dominate at the $\sim$0.1--0.2,dex level \citep{conroy_modeling_2013}.

\section{Conclusion} \label{sec:conclusions}

A systematic comparison of parametric and non-parametric SFH models across \texttt{dynesty}, \texttt{nautilus}, and \texttt{emcee} samplers identified the non-parametric \texttt{emcee} configuration as the preferred description of PEARLSDG. The recovered metallicity and the quenched, mass-concentrated SFH are consistent with those of a typical group galaxy quenched by environmental processes. These results highlight the importance of flexible SFH parameterizations and robust posterior sampling when interpreting dwarf galaxy stellar populations in complex environments.

\begin{acknowledgments}
The authors also thank the MMT Queue Observers for carrying out the 2024 Hectospec observations, and acknowledge K.~Rines for the archival 2007 Hectospec program whose data were incorporated into the combined spectrum analyzed here.

R.A.W.\ acknowledges NASA JWST Interdisciplinary Scientist grants NAG5-12460 and
NNX14AN10G, and STScI grants HST-GO-16252 and JWST-GO-01176 (operated by AURA
under NASA contract NAS5-26555). This research used the NASA/IPAC Extragalactic
Database (NED), funded by NASA and operated by Caltech, the NASA-Sloan Atlas
\citep{blanton_improved_2011}, and \texttt{Astropy} \citep{astropy_2022}.
\end{acknowledgments}

\facilities{JWST (NIRCam), MMT (Hectospec)}
\software{{\sc Prospector} \citep{johnson_stellar_2021}, \texttt{emcee}
\citep{foreman-mackey_emcee_2013}, \texttt{dynesty} \citep{speagle_dynesty_2020},
\texttt{nautilus} \citep{lange_nautilus_2023}, \texttt{Astropy}
\citep{astropy_2022}, \texttt{python-fsps} \citep{conroy_fsps_2009},
\texttt{sedpy} \citep{johnson_sedpy_2021}}

\begin{figure*}
    \centering
    \includegraphics[width=0.80\textwidth]{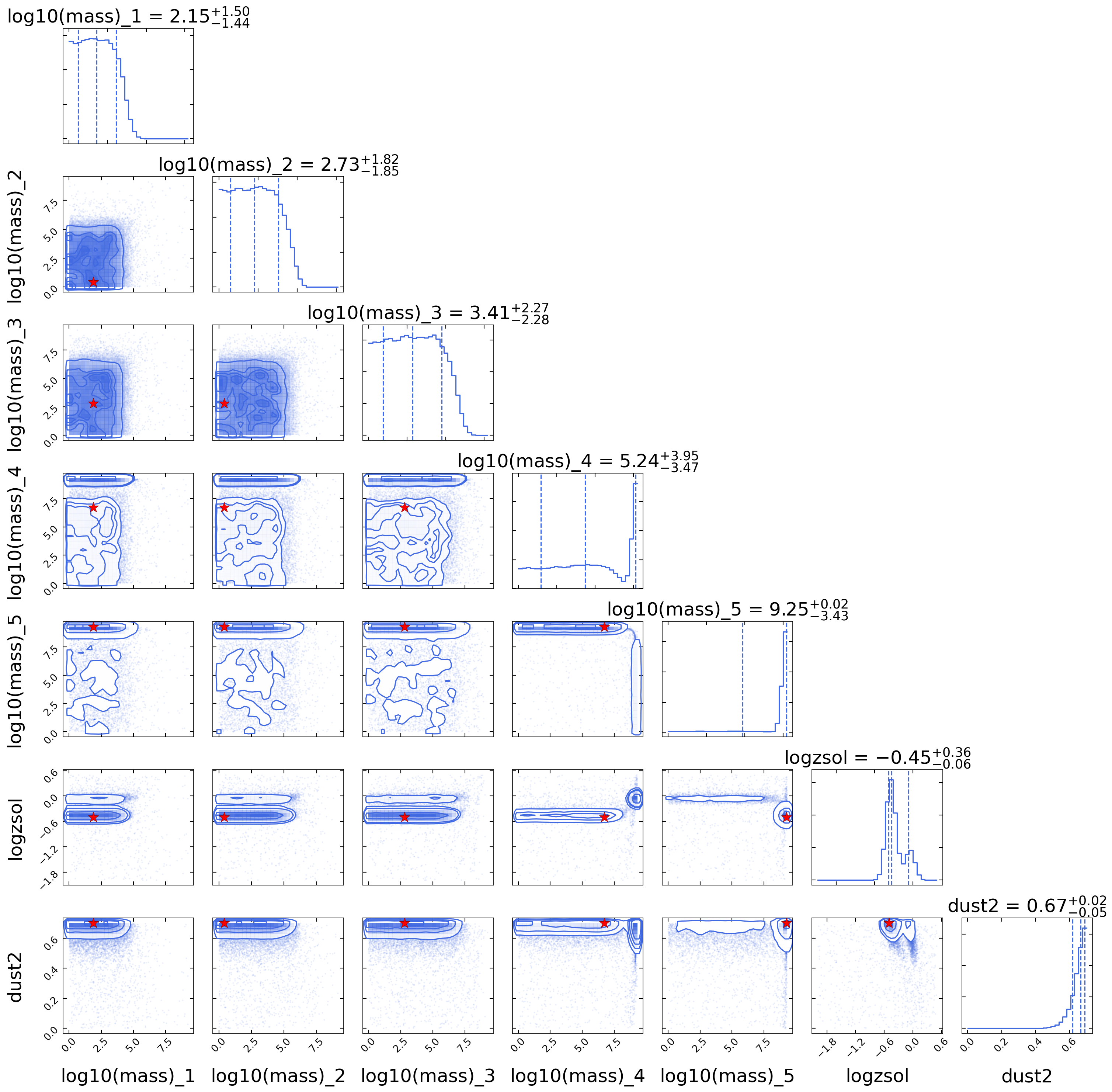}
    \vspace{0.3em}
    \includegraphics[width=0.80\columnwidth]{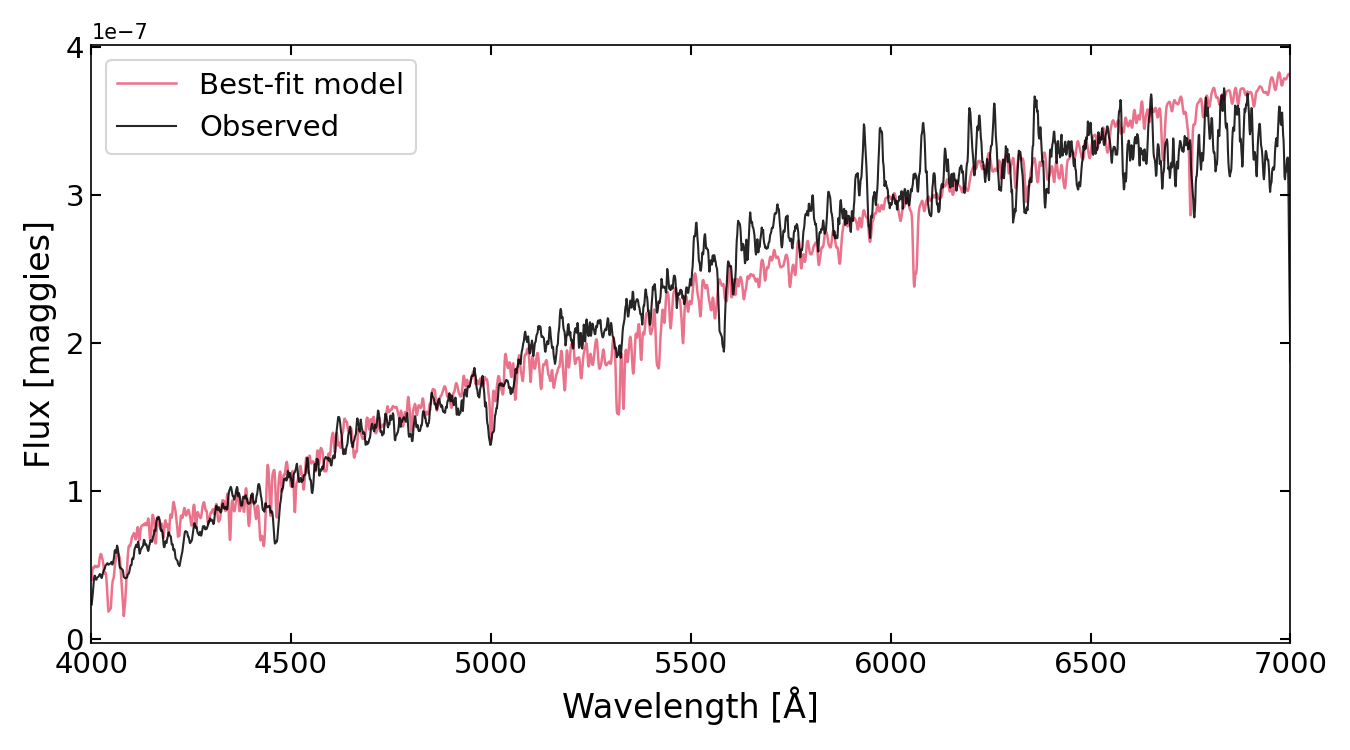}
    \vspace{0.3em}
    \includegraphics[width=0.80\textwidth]{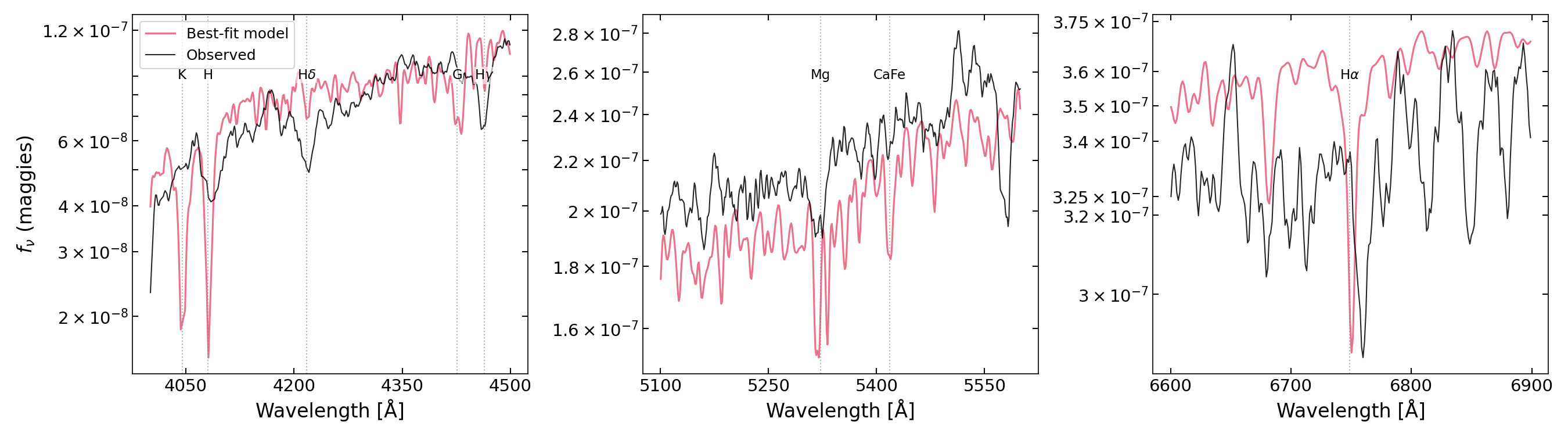}
    \captionsetup{font=footnotesize}
    \caption{
    \textit{Top:} Posterior distributions from the preferred non-parametric
    {\sc emcee} SED fit for the five stellar-mass bins, metallicity, and dust
    attenuation. Red points mark maximum-likelihood estimates.
    \textit{Middle:} Observed Hectospec spectrum (gray) vs.\ best-fit model (red),
    4000--7000\,\AA.
    \textit{Bottom:} Zoomed views around key absorption features as labeled.
    }
    \label{fig:emcee_nonparam}
\end{figure*}

\bibliography{references}
\bibliographystyle{aasjournalv7}

\end{document}